
\input phyzzx
\baselineskip 24pt plus 1pt minus 1pt

\hfill\vbox{\hbox{IC/94/116}
\hbox{IMSC/94-31}
\hbox {August, 1994}}\break
\def\define#1#2\par{\def#1{\Ref#1{#2}\edef#1{\noexpand\refmark{#1}}}}
\def\con#1#2\noc{\let\?=\Ref\let\<=\refmark\let\Ref=\REFS
         \let\refmark=\undefined#1\let\Ref=\REFSCON#2
         \let\Ref=\?\let\refmark=\<\refsend}


\define\VENE
G. Veneziano, Phys. Lett. {\bf B265} (1991) 287;
K. Meissner and G. Veneziano, Phys. Lett. {\bf B267} (1991) 33;
K. Meissner and G. Veneziano, Mod. Phys. Lett. {\bf A6} (1991) 3397.

\define\MAHARANA
M. Gasperini, J. Maharana and G. Veneziano, Phys. Lett. {\bf B272}
(1991) 277.

\define\CALLAN
C. G. Callan, D. Friedan, E. J. Martinec and M. J. Perry, Nucl. Phys.
{\bf B262} (1985) 593; A. Sen, Phys. Rev. Lett. {\bf 55 } (1985) 1846,
Phys. Rev {\bf D32} (1985) 2102.

\define\COOPER
A. Cooper, L. Susskind and L. Thorlacious, Nucl. Phys. {\bf B363}
(1991) 132.

\define\GAN
C. G. Callan and Z. Gan, Nucl. Phys. {\bf B272} (1986) 647;
S. R. Das and B. Sathiapalan, Phys. Rev. Lett. {\bf 56} (1986) 2664;
A. A. Tseytlin, Phys. Lett. {\bf B264} (1991) 311.

\define\KOSTEL
V. Alan Kostelecky and M. J. Perry, Nucl. Phys. {\bf B414} ((1994) 174.

\define\RAITEN
E. Raiten, Nucl. Phys. {\bf B416} (1994) 881.

\define\MULLER
M. Mueller, Nucl. Phys. {\bf B337} (1990) 37.

\define\NAPPI
C. Nappi and E. Witten, Phys. Lett. {\bf B293} (1992) 309.

\define\LUST
C. Kounnas and D. Lust, Phys. Lett. {\bf B289} (1992) 56.

\define\TSEYTLIN
A. A. Tseytlin and C. Vafa, Nucl. Phys. {\bf B372} (1992) 443.

\define\KASNER
E. Kasner, Am. J. Math {\bf 43} (1921) 217.

\define\MISNER
C. W. Misner, Phys. Rev. Lett. {\bf 22} (1969) 1071.

\define\BELINSKY
V. A. Belinsky, I. M. Khalatnikov and E. M. Lifshitz, Adv. in Phys. {\bf
19} (1970) 525.

\define\KIRITSIS
A. A. Tseytlin, Imperial preprint, TP/93-04/36 (hepth/9404191);
E. Kiritsis and C. Kounnas, CERN preprint, CERN-TH 7219/94 (hepth/9404092).

\define\LINDE
A. D. Linde, Phys. Lett. {\bf B108} (1982) 389.

\define\KS
V. A. Kostelecky and S. Samuel, Phys. Rev. {\bf D42} (1990) 1289.

\title{\bf{TACHYON CONDENSATES AND ANISOTROPIC UNIVERSE}}
\author{Swapna Mahapatra\foot{Permanent address: Institute of
Mathematical Sciences, C. I. T. Campus, Madras-600113, India.}
\foot{e-mail: swapna@imsc.ernet.in}\break
and
\break
{}~~Sudipta
Mukherji\foot{e-mail: mukherji@ictp.trieste.it}}
\address{International Centre for theoretical Physics,\break
34100 Trieste, Italy}

\abstract
We investigate the cosmological solutions in closed bosonic string
theory in the presence of non zero tachyon condensate. We specifically
obtain time dependent solutions which describes
an anisotropic universe. We also discuss the nature of such time dependent
solutions when small
tachyon fluctuations
around the condensate are taken into account.
\endpage
String theory is likely to provide us with a consistent quatum theory
of gravity. The issue of understanding the structure of space-time
singularities
is one of the important question and one hopes to have a better
understanding of the nature of such singularities by studying them in the
context of string theory.
As we know, the lowest lying fields in the case of
closed bosonic string theory are the metric, antisymmetric tensor field,
dilaton and a tachyon. In the sigma model approach, the requirement of
vanishing beta function equations for various background fields
ensures conformal invariance of the theory \CALLAN. On the other hand,
these beta function equations can be derived as the target space equations
of motion for the tachyon $T$, dilaton $\Phi$, metric $G_{\mu\nu}$ and the
anti
symmetric tensor field $B_{\mu\nu}$ from the low energy effective action.
One then looks for consistent solutions of the beta function equations
which represent nontrivial backgrounds for string propagation. These
solutions in string theory are qualitatively quite different from their
corresponding counter part in general theory of relativity due to the
presence of a nontrivial dilaton field. String propagation in cosmological
backgrounds have also been discussed before \foot{see for example
\MULLER\VENE\NAPPI\LUST\TSEYTLIN\MAHARANA.}.

The beta function analysis
including the tachyon has been cosidered in ref \GAN.
The existence of the tachyon destabilizes the canonical 26-dimensional
vacuum. But it is quite possible that the collective effects may stabilize
the closed string in another ground state containing a non zero tachyon
expectation value. In \KS, such a possibility was investigated starting from
a covariant string field theory through mass-level truncation scheme.
It was shown that upto cubic order in tachyon field, the string
field theory gets a  contribution to the tachyon potential term of the
form
$$
V(\hat T) = -{2\over{\alpha^\prime}}{\hat T}^2 + {\hat g\over 6}{\hat T}^3.
\eqn\onea
$$
Here $\hat T$ is the tachyon field apearing in the string field theory
and $\hat g$ is the three point tachyon coupling at zero momentum.
{}From \onea , it is
clear that the potential has a minimum at nonzero value of tachyon
along with a maximum at $\hat T = 0$. It is possible that higher order
corrections to \onea~will destroy the structure of the potential, but
nevertheless, it is very tempting to assume that such a non-trivial
minimum exists even when the higher order corrections in the tachyon potential
are taken in to account.

When the tachyon is settled at the minimum of the potential,
the nature of an isotropic universe has been discussed recently in \KOSTEL.
Consequences on the metric for small tachyon fluctuations and possible
cosmological scenario has also been considered \RAITEN.
In this note, we find that such a tachyon condensate does not isolate
an isotropic universe. Infact, there are solutions of string
equations which describes anisotropic metric leading to an anisotropic
evolution of the universe. Such possibility exists even when the tachyon slowly
fluctuates around its non zero vacuum expectation value.

We start with Bianchi-I type of anisotropic metric
with the line element given by
$$
d s^2 = - d t^2 + \sum_ia_i^2(t)~d x_i^2\eqn\one
$$
where, $i$ = 1,...., $d - 1$, $a_i(t)$'s are the scale factors  in the presence
of non-zero tachyon condensate.

The motivation to consider the anisotropic cosmological model is due to lack
of adequate explanation for the high degree of isotropy in the universe
as observed by experiments. There are arguments given previously that
adiabatic cooling and viscous disssipation might be the reason for
destroying the anisotropy of the universe. Also the process of quantum
particle creation in the anisotropic expansion might have caused a stronger
damping of the anisotropy in the early universe. Though the Friedman-
Robertson-Walker cosmological model is satisfying enough from the point
of view of experiment, however it does not give a satisfactory answer that
why is the universe so homogeneous and isotropic on large scales. So
one needs to go beyond the isotropic case to understand this phenomena.
Our aim is to see if such an anisotropic model is allowed by string theory.
Given the ansatz for the metric, we look for solutions of the beta function
equations for $G_{\mu\nu}$, $\Phi$ and $T$. We assume that the dilaton
and tachyon fields are functions of time only and we consider the $d$
dimensional space time for our analysis.

The action for a string  propagating in a background of graviton, dilaton
and tachyon condensates is given by,

$$
S = -{1\over 2\pi \alpha'}\int d^2 \sigma \sqrt g ~\lbrack g^{a b}
{\partial}_a
X^{\mu}
{\partial}_b X^{\nu} G_{\mu\nu} + ~{1\over 2} \alpha' T
- {1\over 4} \alpha' R^{(2)} \Phi ~\rbrack\eqn\two
$$

Here $g_{a b}$ is the two dimensional metric in the world sheet and
$R^{(2)}$
is the world sheet curvature scalar. The volume two form on the world sheet
has been neglected here, but in principle we could have chosen a nonzero
$B_{\mu\nu}$ field. The background fields are the metric $G_{\mu\nu}$,
dilaton $\Phi$ and the tachyon $T$ and they basically act like couplings in
the corresponding non linear sigma model. The one loop beta function equations
for the metric, tachyon and dilaton are respectively given by,

$$\eqalign{&R_{\mu\nu} = \nabla_{\mu} \nabla_{\nu}\Phi + \nabla_{\mu} T
\nabla_{\nu} T ;\cr
&\nabla^2 T + \nabla_{\mu} \Phi \nabla^{\nu} T = V'(T) ;\cr
&\hat c = R -  (\nabla\Phi)^2 - 2 \nabla^2 \Phi - (\nabla T)^2 -
2 V(T).\cr}\eqn\thre$$

where, $\hat c = {2(d - 26)\over 3\alpha'}$, $V'(T) =
{\partial V(T)\over
\partial T}$ and $V(T)$ is the tachyon potential. As discussed before, these
background field equations can be derived from the target space effective
action in $d$ dimensions, which is given by,

$$
S_{eff} = -{1\over 2 \kappa^2} \int d^d x \sqrt G e^{\Phi} \lbrack
\hat c - R - (\nabla\Phi)^2 + (\nabla T)^2 + 2 V(T) \rbrack\eqn\for
$$

where, $\kappa^2 = 8 \pi G_N$ when $d = 4$. This effective action has been
written in the string frame, where $e^{{\Phi\over 2}}$ serves as the string
coupling constant. This action can be represented in the so called Einstein
frame through the following transformation,

$$
G_{\mu\nu}^E = e^{- {2 \Phi\over d-2}} \,G_{\mu\nu}^{\sigma}\eqn\fiv
$$

Having given the frame work, let us look at the various beta function
equations corresponding to the ansatz for the metric describing the anisotropic
universe. The equations for the $tt$ and $ii$ components of the metric
are respectively given by,

$$
\eqalign{&({\ddot a_1\over a_1} + {\ddot a_2\over a_2} +
{\ddot a_{d -1}\over a_{d-1}})
+ \ddot \Phi(t) - (\dot T)^2 = 0;\cr
&a_i \ddot a_i + a_i \dot a_i \sum_{j \neq i}({\dot a_j\over a_j}) +
a_i \dot a_i \dot \Phi = 0\cr}\eqn\six
$$

The $ti$ component of the metric equation vanishes identically. Tachyon
equation is given by,

$$
\ddot T + \dot T ({\dot a_1\over a_1} + {\dot a_2\over a_2} + ......+
{\dot a_{d- 1}\over a_{d-1}}) + \dot \Phi \dot T + V'(T) = 0
\eqn\sevn
$$

where prime denotes differentiation w.r.t. $T$ and dot denotes derivatives
w.r.t. time. Finally the dilaton equation is given by,

$$
\hat c = (\dot \Phi)^2 + \ddot \Phi + \dot \Phi ({\dot a_1\over a_1} +
{\dot a_2\over a_2} + ......+ {\dot a_{d-1}\over a_{d-1}}) - 2 V(T)\eqn\eit
$$

As we discusssed before, at the extremum of the tachyon potential ($V'(T)
= 0$), $T = T_0$ (constant) can also be a consistent solution apart from the
case of $T = 0$. We note that, for $T = T_0$, $V(T_0) \neq 0$ and $V''(T_0)$
is positive which ensures a stable solution.
We define the Hubble parameter $H_i$ as, $H_i = {\dot a_i\over a_i}$ and
$h_i$'s  are defined as $h_i(t) = ln(\sqrt\beta{\dot a_i\over a_i})$ , where
$\beta$ is a positive number. Eliminating $\Phi$ from equation eqn.$\six$,
we find that they  can be written in a more compact form (in
terms of $h_i$'s) namely,

$$
\beta \ddot h_i = \sum^{d-1}_{i=1} e^{2 h_i}\eqn\nin
$$

This equation can be integrated to give,

$$
(\dot h_i)^2 = {1\over \beta} \sum^{d-1}_{i=1} e^{2 h_i} + k\eqn\ten
$$
where, $k$ is an integration constant and $\dot h_1 =
\dot h_2 =.....\dot h_{d-1}$
(assuming that one finds a consistent solution for $a_i(t)$'s, which
satisfy the above equality). In fact we find that this is true.
%
The dilaton can be written in terms of $h_i$'s as
$$
\dot \Phi = - \dot h_i - {1\over \sqrt\beta}\sum_{i = 1}^{d -1}
e^{h_i}\eqn\twel
$$
We now look for solutions of \ten~for various values of $k$, namely
$k = 0, k > 0$ and $k < 0$.
The simplest case is for $k = 0$. We find that the consistent solutions
for the scale factors and dilaton are given by,
$$
a_1(t) = a_1^0 \,t^{p_1};\quad a_2(t) = a_2^0 \,t^{p_2};\quad .....,
a_{d-1}(t) = a_{d-1}^0 \,t^{p_{d-1}}\eqn\thirt
$$
with the constraint,
$$
\sum^{d-1}_{i=1}(p_i)^2 = 1\eqn\fort
$$
The dilaton is given by,
$$
\Phi(t) = \lbrack 1 - (p_1 + p_2 +......+ p_{d-1})\rbrack\,
{\rm log}({t\over\sqrt\beta}) + \Phi_0\eqn\fiftn
$$
where, $a_1^0$, $a_2^0$,...... and $\Phi_0$ are constants, $p_i$'s are
real numbers. Substituting this into dilaton equation, we find that
$\hat c + 2 V(T_0) = 0$. Note that, the above choice
of $a_i$'s leads to the most general solution and the condition $\dot h_1 =
\dot h_2 = .......= \dot h_{d-1} = \dot h$ is also satisfied as it should, in
order to have a consistant solution of \twel .
The tachyon beta function equation is also satisfied for the choice
$T = T_0 \neq 0$.

Next, we consider the case of positive $k$ ($k>0$). The solutions for
$a_i$'s are given by,

$$
\eqalign{ a_1(t)& = a_1^0 \,{\lbrack {\rm tanh}({\sqrt k\over 2} t)\rbrack}^{
p_1};
\qquad a_2(t) = a_2^0 \,{\lbrack {\rm tanh}({\sqrt k\over 2} t)\rbrack}^{p_2};
\cr
&\qquad ...... a_{d-1}(t) = a_{d-1}^0 \,{\lbrack {\rm tanh}({\sqrt k\over 2} t)
\rbrack}^{p_{d-1}}\cr}\eqn\sixtn
$$

Then for this choice of $a_i(t)$'s, the dilaton is found to be,
$$
\eqalign{\Phi(t) &= \Phi_0 + \lbrack 1 + \sum_{i = 1}^{d -1} p_i \rbrack \,{\rm
log}
{\rm cosh}({\sqrt k\over 2} t) \cr
&+ \lbrack 1 - \sum_{i = 1}^{d-1} p_i \rbrack \,{\rm log} {\rm sinh}({\sqrt
k\over 2}
t) \cr}\eqn\sevtn
$$
As before,
$p_i$'s satisfy the condition,
$\sum_{i = 1}^{d-1} (p_i)^2 = 1$.
Substituting the above solutions in the dilaton beta function equation,
we find,
$$
\hat c + 2 V(T_0) = k\eqn\nintn
$$

Next, consider the case when $k$ is negative ($k<0$). In this case,
the solutions for the scale factors are given by,

$$
\eqalign{a_1(t) &= a_1^0~{\lbrack {\rm tan}({\sqrt k\over 2} t)\rbrack}^{p_1};
\qquad a_2(t) = a_2^0~{\lbrack {\rm tan}({\sqrt k\over 2} t)\rbrack}^{p_2};\cr
& .......a_{d -1}(t) = a_{d -1}^0~{\lbrack {\rm tan}({\sqrt k\over 2} t)
\rbrack}^{p_{d -1}}\cr}\eqn\twenty
$$

The solution for dilaton is given by,

$$
\eqalign{\Phi & = \Phi_0 + \lbrack 1 - \sum_{i = 1}^{d -1} p_i \rbrack \,
{\rm log}{\rm cos}({\sqrt k\over
2} t) \cr &+ \lbrack 1 - \sum_{i =1}^{d -1} p_i \rbrack \,{\rm log} {\rm
sin}({\sqrt k\over 2} t)\cr}
\eqn\twentone
$$
with the conditon, $\sum_{i = 1}^{d - 1}(p_i)^2 = 1$. The dilaton beta
function equation in this case implies that, $\hat c + 2 V(T_0) = - k$.

We notice that for small $t$, the behaviour of the solutions for the $k>0$
and $k<0$ cases are the same as that of $k = 0$ case. As $t$ increases,
the scale factors tend towards constant values namely $a_1^0$, $a_2^0$,
.... etc. (implying asymptotically flat space) and dilaton grows
linearly with time.  For $k<0$, we find that the scale factors diverge
for $t \rightarrow {\pi\over \sqrt k}$, which implies that the universe
expands to infinite size in finite amount of time.
When all the $p_i$ are equal, our solution  reduces to that of the
isotropic model considered by Kostelecky and Perry \KOSTEL.

We also notice that the Kasner model for the
anisotropic universe \KASNER\ is a solution of the string beta
function equations with a constant dilaton background. Kasner model
in four dimensions is described by the metric,

$$
d s^2 = - d t^2 + t^{2 p_1} d x^2 + t^{2 p_2} d y^2 + t^{2 p_3} d z^2
\eqn\twentthr
$$
In this metric, the $p_i$'s are real numbers, which satisfy the relation,
$$
\sum_{i =1}^3 p_i = 1; \qquad \sum_{i = 1}^3 (p_i)^2 = 1\eqn\twentfour
$$

Note that, our most general solution also required the constraint
$\sum_{i = 1}^3 (p_i)^2 = 1$ in four dimensions. The constraint
$\sum_{i = 1}^3 p_i = 1$, automatically implies a constant dilaton
in our case. Here one of the $p_i$'s has to be negative to satisfy
the constraint $\sum_i p_i = 1$ in the Kasner model. The limits are
found to be,
$$
-{1\over 3} \leq p_1\leq 0; \qquad 0\leq p_2\leq {2\over 3};
\qquad {2\over 3}\leq p_3\leq 1.\eqn\twentfv
$$

More generally, in the mixmaster universe (which was originally
introduced to explain the inhomogeneous behaviour of the universe near
a singularity) \MISNER, the Kasner exponents $p_i$'s can become
functions of time near the singularity. The Kasner like behaviour
exists for the exponents corresponding to some fixed $u$  parameter.
They are given by,
$$
p_1(u) = - {u\over {1 + u + u^2}}; \qquad p_2(u) = {1 + u
\over {1 + u + u^2}}; \qquad p_3(u) = {u(1 + u)\over {1 + u + u^2}}
\eqn\twentsix
$$
with the constraint, $\sum_{i = 1}^3 (p_i)^2 = 1$; $\sum_{i=1}^3
p_i = 1$. In the string theory frame work, the mixmaster metric is
also found to be a consistent solution with a constant dilaton
background. Remember that, we are doing all these analysis in a
background where tachyon has a constant expectation value.

Finally, we consider small tachyon fluctuation
around its background value and investigate
the possibility of a viable inflationary scenario in the case of
anistropic universe. The tachyon potential given in \onea~appears
in non-polynomial string field theory. It is known that various
fields that appear in this field theory and the corresponding
sigma model coupling constants (or  the fields in the low energy effective
field theory derived from it) are related through complicated field
redefinitions. But as far as the tachyon field in two theories are
concerned, they are related simply by \KOSTEL
$$
T = {\hat g\over 4}\alpha^\prime \hat T,~~~V(T) = {{\hat g}^2\over {16}}
{\alpha^\prime}^2 {\hat V}(\hat T)
$$
so that,
$$
V(T) = - {2\over \alpha'} T^2 + {2\over {3\alpha^\prime}} T^3
\eqn\twentsev
$$

The inflationary scenario \LINDE\ involving such a tachyon effective
potential has been discussed recently in ref. \RAITEN. They ofcourse
assumed a $k = 0$ Friedman-Robertson-Walker universe to start with.
Here we will consider the anisotropic model as before. The tachyon
plays the role of the scalar field in inflationary model, which undergoes
a slow roll during the inflation. Since here we consider fluctuations
of the tachyon around the condensate,
we expect the solutions of the beta function
equations to change, though we can not solve these equations exactly
as we could do before. We rewrite the beta function equations in terms
of $h_i$'s. Eliminating $\Phi$ from the beta function equations, we get,
$$
(\dot T)^2 = \ddot h_i - {1\over \beta}
\sum_{i = 1}^{d-1} e^{2 h_i},\eqn\twenteit
$$
$$
\ddot T - \dot h \dot T + V'(T) = 0\eqn\twentnin
$$
Note that the l.h.s. of eqn.\twenteit~
was zero in the constant tachyon background
case and equation \twentnin~ was identically zero as $T = T_0$ was the true
minimum. We have written eqn. \twenteit~ assuming again that one finds the
consistent solutions $h_i$'s, where all the $h_i$'s differ by constants
only, which we have noticed in the previous case. We assume that as
time $t$ increases, the tachyon rolls towards the true minimum at $T =
T_0$. This means that $(\dot T)^2$ and $\ddot T$ terms are small in the
above consideration. We take $T$ as  a linear function of $t$,
approximated by,

$$
T = C t + T_0\eqn\thirt
$$
where, $C$ and $T_0$ are constants. $\ddot T$ term then drops out from
equation \twentnin~ given above and
we have the following two equations to solve,
$$
C \dot h_i - V'(T) = 0\eqn\thirton
$$
$$
(\dot h_i)^2 - {1\over\beta} \sum_{i = 1}^{d - 1} e^{2 h_i} - 2 C^2 h_i(t) =
k'\eqn
\thirttwo
$$
where, $k'$ is an integration constant. We have used the fact that
all the $\dot h_i$'s are equal. It is difficult to find exact
solutions of these equations. So we retain only upto order $h_i$
terms in the exponential as has been done in ref. \RAITEN. Then we have
the following equation to solve,
$$
(\dot h_i)^2 - {1\over \beta}\lbrack (d - 1) + 2 \sum_{i = 1}^{d -1}
h_i\rbrack - 2 C^2 h_i = k'\eqn\thirtthree
$$
Since all the $h_i$'s differ by constants only, we can write eqn. \thirtthree
as,
$$
(\dot h)^2 - {(d -1)\over\beta} \lbrack 1 + 2 h(t) \rbrack -
2 C^2 h(t) = k' + k''\eqn\thirtfor
$$
where, $k''$ is the corresponding constant factor which we generate
by writing $h_i(t) = h(t) + k_i$. $h(t)$ is nothing but the
$t$ dependent part of ${\rm log}(\sqrt \beta{\dot a_i\over a_i})$,
which we take to be equal for all the $a_i$'s. We can rewrite \thirtfor
as,

$$
(\dot h)^2 - C_1 h(t) - C_2 = 0\eqn\thirfuv
$$
where,
$$
C_1 = {2(d - 1)\over\beta} + 2 C^2; \qquad C_2 = {d - 1\over\beta} +
k' + k''\eqn\thirsix
$$
Integrating \thirfuv, we obtain,
$$
h(t) = h_0 \pm t \sqrt{C_1 h_0 + C_2} + {1\over 4} C_1 t^2\eqn\thirsevn
$$
And the solutions for $h_i$'s are given by,
$$
h_i(t) = (h_0 + k_i) \pm t \sqrt{C_1 h_0 + C_2} + {1\over 4} C_1
t^2\eqn\thiorteit
$$
Even though $h_i$s differ from each other only by a constant, the
correspoding $a_i$s can differ from each other in a time dependent
manner. As a result it would generically lead to anisotropic universe.

In this letter, we have found consistent solutions of beta function
equations which describes an anisotropic universe in the presence of nonzero
tachyon condensate.
We have also investigated the possibility of
obtaining such solutions for the metric and dilaton for small tachyon
fluctuations around its background value. This looks like
a viable candidate for studying the inflationary cosmology in string
theory context. Recently, time dependent solutions in string theory have
been considered in ref. \KIRITSIS, where the modulus field changes with time,
thereby inducing a dynamical topology change in the theory. The time de
pendent radii there are analogus to the scale factors $a_i(t)$. Our
solutions for the scale factors for the cases
$k = 0$, $k< 0$, $k > 0$ are similar to those in \KIRITSIS. One has to
investigate in
detail the tachyon induced inflation in an isotropic universe. It would also
be interesting to find out an exact solution in the above anisotropic
case, having a conformal field theory description of the
underlying sigma model action.

\noindent {\bf Acknowledgements:} We would like to thank Professor
Abdus Salam, the International Atomic Energy Agency and UNESCO for
hospitality at the International Centre for Theoretical Physics, Trieste,
where this work was done.

\refout
\end